\documentclass[preprint,showpacs,preprintnumbers,amsmath,amssymb]{revtex4}
\usepackage{epsfig}
\usepackage{graphicx}
\usepackage{dcolumn}
\usepackage{bm}

\def\beq{\begin{equation}}
\def\eeq{\end{equation}}
\def\eeqn{\end{equation}}
\newcommand\iden{\leavevmode\hbox{\small1\normalsize\kern-.33em1}}

\newcommand{\bea} {\begin{eqnarray}}
\newcommand{\eea} {\end{eqnarray}}
\newcommand{\be}{\begin{equation}}
\newcommand{\ee}{\end{equation}}

\newcommand{\ba}{\begin{array}}
\newcommand{\ea}{\end{array}}

\let\jnfont=\rm
\def\NPB#1,{{\jnfont Nucl.\ Phys.\ B }{\bf #1},}
\def\PLB#1,{{\jnfont Phys.\ Lett.\ B }{\bf #1},}
\def\EPJC#1,{{\jnfont Eur.\ Phys.\ Jour.\ C }{\bf #1},}
\def\PRD#1,{{\jnfont Phys.\ Rev.\ D }{\bf #1},}
\def\PRL#1,{{\jnfont Phys.\ Rev.\ Lett.\ }{\bf #1},}
\def\MPLA#1,{{\jnfont Mod.\ Phys.\ Lett.\ A }{\bf #1},}
\def\JPG#1,{{\jnfont J.\ Phys.\ G }{\bf #1},}
\def\CTP#1,{{\jnfont Commun.\ Theor.\ Phys.\ }{\bf #1},}
\def\JHEP#1,{{\jnfont JHEP \ }{\bf #1},}
\def\NPPS#1,{{\jnfont Nucl.\ Phys.\ Proc.\ Suppl.\ }{\bf #1},}
\def\CPC#1,{{\jnfont Computl.\ Phys.\ Commun.\ }{\bf #1},}
\def\CPL#1,{{\jnfont Chin.\ Phys.\ Lett. }{\bf #1},}
\def\APPB#1,{{\jnfont Acta\ Phys.\ Polon.\ B }{\bf #1},}

\def\lsim{\raise0.3ex\hbox{$<$\kern-0.75em\raise-1.1ex\hbox{$\sim$}}}
\def\gsim{\raise0.3ex\hbox{$>$\kern-0.75em\raise-1.1ex\hbox{$\sim$}}}

\begin{document}

\title{\ \\[10mm] LHC diphoton and Z+photon Higgs signals in the Higgs triplet model with Y=0}

\author{Lei Wang, Xiao-Fang Han}


\affiliation{ Department of Physics, Yantai University, Yantai
264005, China}


\begin{abstract}
We study the implications of the LHC diphoton and Z+photon Higgs
signals on the Higgs triplet model with Y=0, which predicts two
neutral CP-even Higgs bosons $h$, $H$ and a pair of charged Higgs
$H^\pm$. We discuss three different scenarios: (i) the observed
boson is the light Higgs boson $h$; (ii) it is the heavy Higgs boson
$H$; (iii) the observed signal is from the almost degenerate $h$ and
$H$. We find that the inclusive Higgs diphoton rates in the first
two scenarios can be enhanced or suppressed compared to the SM
value, which can respectively fit the ATLAS and CMS diphoton data
within $1\sigma$ range. The inclusive $ZZ^*$ rates are suppressed,
which are outside $1\sigma$ range of ATLAS data  and within
$1\sigma$ range of CMS data. Meanwhile, another CP-even Higgs boson
production rate can be suppressed enough not to be observed at the
collider. For the third scenario, the Higgs diphoton rate is
suppressed, which is outside $1\sigma$ range of ATLAS data, and the
$ZZ^*$ rate equals to SM value approximately. In addition, we find
that the two rates of $h\to \gamma\gamma$ and $h\to Z\gamma$ have
the positive correlations for the three scenarios.
\end{abstract}



\maketitle

\section{Introduction}
The CMS and ATLAS collaborations have announced the observation of a
new boson around 125.5 GeV \cite{cmsh,atlh}, which is corroborated
by the Tevatron search results \cite{12073698:4}. The properties of
this particle with large experimental uncertainties are consistent
with the SM Higgs boson. Among the various signals, the diphoton and
$ZZ^*$ are the cleanest channels of seraching for the Higgs boson.
The CMS and ATLAS have presented the constraints
\cite{cmsrr,atlasrr},\bea && R_{\gamma\gamma}=0.77\pm
0.27,~~~~~R_{ZZ^*}=0.92\pm0.28~~\text{(CMS)},~~~~~\nonumber
\\&&R_{\gamma\gamma}=1.6\pm0.3, \hspace{1.08cm} R_{ZZ^*}=1.5\pm0.4~~\text{(ATLAS)}.\eea The CMS
collaboration has released their results of the measurement of
$Z\gamma$ and set an upper limit on the ratio $R_{Z\gamma} < 10$
\cite{zrdata}.

The recent Higgs data has been discussed in the SUSY
models\cite{susy}, little Higgs models \cite{hrrlh} and the
extensions of Higgs field models, such as the two-Higgs-doublet
model \cite{2hdm}, the Higgs triplet model (Y=2) \cite{triplet}, the
models with septuplet \cite{wcao} and color-octet scalar
\cite{octet}. In this work, we will study the implications of the
LHC diphoton and $Z$+photon Higgs signals on the Higgs triplet model
with Y=0 (HTM0) \cite{htm0}, which predicts two neutral CP-even
Higgs bosons $h$, $H$ and a pair of charged Higgs $H^\pm$. We will
discuss three different scenarios: (i) the observed boson is the
light Higgs $h$, and the heavy Higgs $H$ is not observed at the LHC;
(ii) it is the heavy Higgs $H$, and the light Higgs $h$ is not
observed at the LEP; (iii) the observed signal is from the almost
degenerate $h$ and $H$. Also we will pay the particular attention to
the correlations between $h\to Z\gamma$ and $h\to \gamma\gamma$.
Since both of the rates are loop-induced by charged particles, they
should be closely correlated. Any new physics effects manifested in
the diphoton decay should also alter the $Z\gamma$ decay
\cite{hzr,geng}

Our work is organized as follows. In Sec. II we recapitulate the
Higgs triplet model with Y=0. In Sec. III we discuss the LHC
diphoton  Higgs signal and the correlations between $h\to Z\gamma$
and $h\to \gamma\gamma$. Finally, we give our conclusion in Sec. IV.

\section{Higgs triplet model with Y=0}

In the HTM0, a real $\rm{SU(2)_L}$ triplet scalar field $\Sigma$
with $Y = 0$ is added to the SM Lagrangian in addition to the
doublet field $\Phi$. These fields can be written as
\begin{eqnarray} \Sigma &=\frac{1}{2}\left(
\begin{array}{cc}
\delta^0 & \sqrt{2}\delta^{+} \\
\sqrt{2}\delta^{-} & -\delta^0\\
\end{array}
\right),  \qquad \Phi=\left(
                    \begin{array}{c}
                      \phi^+ \\
                      \phi^0 \\
                    \end{array}
                  \right).
\end{eqnarray}
 The renormalizable scalar potential can
be written as \cite{kwang} \bea V(\Phi,\Sigma) & = & - \mu^2 \
\Phi^\dagger \Phi \ + \ \lambda_0  \ \left( \Phi^\dagger \Phi
\right)^2 \ - \frac{1}{2}\ M^2_{\Sigma} F \ + \ \frac{b_4}{4} F^2 \
+ \ a_1 \ \Phi^\dagger \Sigma \Phi +\frac{a_2}{2} \Phi^\dag \Phi F \
, \label{potent} \eea where $F\equiv \left(\delta^0\right)^2 +
2\delta^+\delta^-$ and all the parameters are real. The Higgs
doublet and triplet fields can acquire vacuum expectation values
\begin{equation}
\langle \Phi \rangle = \frac{1}{\sqrt{2}} \left(
                    \begin{array}{c}
                      0 \\
                      v_d \\
                    \end{array}
                  \right), \qquad \langle \Delta \rangle = \frac{1}{2}
\left(
\begin{array}{cc}
v_t & 0 \\
0 & -v_t\\
\end{array}
\right) \label{vacuum}
\end{equation} with $v^2=v_d^2+4v_t^2\approx(246~\rm{GeV})^2$.

After the spontaneous symmetry breaking, the Lagrangian of Eq.
(\ref{potent}) predicts the four physical Higgs bosons, including
two CP-even Higgs bosons $h$, $H$ and a pair of charged Higgs
$H^\pm$. These mass eigenstates are in general mixtures of the
doublet and triplet fields. The mass matrixes of neutral and charged
Higgs bosons are \cite{kwang}
\begin{equation}
 {\cal M}_{0}^2 = \left( \begin{array} {cc}
2 \lambda_0 v_d^2  &  - a_1 v_d / 2 + a_2  v_d  v_t \\
- a_1  v_d/ 2  +  a_2  v_d  v_t  & 2 b_4  v_t^2 + \frac{a_1 v_d^2}{4
v_t}
\end{array} \right)\equiv\left( \begin{array} {cc}
A  &  B \\
B  & C
\end{array} \right), \\
{\cal M}_{\pm}^2 = \left( \begin{array} {cc}
a_1 v_t  &  a_1 v_d / 2  \\
a_1  v_d/ 2 & \frac{a_1 v_d^2}{4 v_t}
\end{array} \right).
\label{matrix}\end{equation} The physical mass eigenstates and the
unphysical electroweak eigenstates are related by rotations through
two mixing angles $\theta_0$ and $\theta_+$:
\begin{eqnarray}
\label{eq:neutralmix}
\left( \begin{array}{c} h \\ H\end{array} \right) & = &
\left(\begin{array}{ccc} \cos \theta_0 & \sin \theta_0 \\ - \sin \theta_0 & \cos \theta_0\end{array} \right)
\left( \begin{array}{c} \phi^0 \\ \delta^0 \end{array} \right) \ , \\
& & \nonumber \\
\label{eq:chargemix} \left( \begin{array}{c} H^\pm \\ G^\pm
\end{array} \right) & = & \left(\begin{array}{cc} -\sin \theta_\pm &
\cos \theta_\pm  \\ \cos\theta_\pm & \sin \theta_\pm  \end{array}
\right) \left( \begin{array}{c} \phi^\pm \\ \delta^\pm \end{array}
\right) \ .
\end{eqnarray}
Where the Goldstone boson $G^\pm$ is eaten by the gauge bosons.

Since the experimental value of the $\rho$ parameter is near unity
\cite{ro}, $4v^2_t/v^2_d$ is required to be much smaller than unity.
In our calculation, $v_t$ is taken as 1 GeV. The mixing angle
$\theta_\pm$ is proportional to $\frac{v_t}{v_d}$, therefore it is
very small. The charged Higgs mass is given as \be M_{H^{\pm}}^2 =
a_1 v_t \left( 1 + \frac{ v_d^2}{4 v_t^2} \right).\label{mcharge}\ee

The neutral mixing angle $\theta_0$ is given as \bea c_{0}\equiv
\cos\theta_0&=&\frac{1}{\sqrt{2}}\left(1-\frac{A-C}{\sqrt{(A-C)^2+4B^2}}\right)^{1/2},\nonumber\\
s_{0}\equiv \sin\theta_0&=&-\frac{1}{\sqrt{2}}\frac{B}{\mid
B\mid}\left(1+\frac{A-C}{\sqrt{(A-C)^2+4B^2}}\right)^{1/2}.\label{hHangle}\eea
Where\beq c_0>\frac{1}{\sqrt{2}}~ \text{for}~ C>A,~~~
c_0<\frac{1}{\sqrt{2}}~ \text{for}~~~
C<A,~~~c_0\rightarrow\frac{1}{\sqrt{2}}~ \text{for}~ C\rightarrow
A.\label{CA}\eeq

The neutral Higgs boson masses are given as \bea
m^2_h&=&\frac{1}{2}\left(A+C-\sqrt{(A-C)^2+4B^2}\right),\nonumber\\
m^2_H&=&\frac{1}{2}\left(A+C+\sqrt{(A-C)^2+4B^2}\right)\label{hHmass}.\eea

In our calculations, the involved Higgs couplings are listed as
\cite{kwang}\bea hf\bar{f}&:&-i\frac{m_f}{v_d}c_0,\hspace{2.8cm}
Hf\bar{f}:i\frac{m_f}{v_d}s_0,\nonumber\\
ZZh&:&\frac{2im_Z^2}{v_d}c_0g^{\mu\nu},\hspace{2.9cm}
ZZH:-\frac{2im_Z^2}{v_d}s_0g^{\mu\nu},\nonumber\\
W^+W^-h&:&ig^2_2\big(\frac{1}{2}v_dc_0+2v_ts_0\big)g^{\mu\nu},\hspace{1.0cm}
W^+W^-H:ig^2_2\big(-\frac{1}{2}v_ds_0+2v_tc_0\big)g^{\mu\nu},\nonumber\\
\gamma H^+H^-&:&ie\,\big(p'-p\big)^\mu,\hspace{3.3cm}
ZH^+H^-: i\big(g_2c_W-\frac{m_Z}{v_d}s_+^2)\big(p'-p\big)^\mu,\nonumber\\
H^+H^-h&:&-i\big(a_1c_+s_+c_0-\frac{1}{2}a_1s_+^2s_0+a_2v_dc_
+^2c_0+a_2v_ts_+^2s_0+2b_4v_tc_+^2s_0+2\lambda_0v_ds_+^2c_0\big),\nonumber\\
H^+H^-H&:&-i\big(-a_1c_+s_+s_0-\frac{1}{2}a_1s_+^2c_0-a_2v_dc_
+^2s_0+a_2v_ts_+^2c_0+2b_4v_tc_+^2c_0-2\lambda_0v_ds_+^2s_0\big).\nonumber\\
\label{coupling}\eea Where $s_+=\sin\theta_+$ and
$c_+=\cos\theta_+$. All the momenta flow into the vertex.

\section{The Higgs diphoton and $Z\gamma$ rates at the LHC}
In our calculations, we take $m_h$, $m_H$, $a_2$, $b_4$ and $v_d$,
$v_t$ as the input parameters, which can determine the values of
$\lambda_0$, $a_1$, $m_{H^\pm}$. As mentioned above, $v_t$ is taken
as 1 GeV. The perturbativity can give the strong constraints on
$a_2$ and $b_4$,\beq -2\sqrt{\pi}\leq
a_2\leq2\sqrt{\pi},~~~~~-2\sqrt{\pi}\leq b_4\leq2\sqrt{\pi}.\eeq The
electroweak $T$ parameter can give the constraints on the splitting
of $m_H$ and $m_{H^{\pm}}$, $(m_H-m_{H^{\pm}})^2 < 0.96~m_W^2$
\cite{kwang}. Since the coupling $H^\pm \bar{f}_i f_j$ is sizably
suppressed by $s_+$, the search experiments through the top quark
decay hardly give the constraints on $H^\pm$. The experimental data
at the LEP gives the lower bound of the charged Higgs mass,
$m_{H^{\pm}}>$ 79.3 GeV \cite{mcmass}.

We discuss three different scenarios: (I) the observed boson is the
light Higgs $h$, $m_h=125.5$ GeV and 135 GeV $\leq m_H\leq500$ GeV;
(II) it is the heavy Higgs $H$, $m_H=125.5$ GeV and 80 GeV $\leq
m_h\leq110$ GeV; (III) the observed signal is from the almost
degenerate $h$ and $H$, $m_h\simeq m_H\simeq125.5$ GeV.

As shown in the Eq. (\ref{coupling}), the $h$ couplings to
$f\bar{f}$ and $WW$ are proportional to $c_0$ while these couplings
of $H$ are proportional to $s_0$. Due to $v_t\ll v_d$ and
$s_+\rightarrow 0$, the $h$ couplings to $WW$ and $H^+H^-$ are
sensitive to $c_0$ while these couplings of $H$ are sensitive to
$s_0$. Therefore, the cross sections and the decay widths of $h(H)$
normalized to SM values can be given as
\bea&&\frac{\sigma\left(~gg\to h(H)~\right)}{\sigma_{SM}\left(~gg\to
h(H)~\right)}\simeq\frac{\sigma\left(~pp\to
jjh(H)~\right)}{\sigma_{SM}\left(~pp\to
jjh(H)~\right)}\nonumber\\&&\simeq\frac{\sigma\left(~pp\to
Vh(H)~\right)}{\sigma_{SM}\left(~pp\to
Vh(H)~\right)}\simeq\frac{\sigma\left(~pp\to
h(H)t\bar{t}~\right)}{\sigma_{SM}\left(~pp\to
h(H)t\bar{t}~\right)}\simeq c_0^2 ( s_0^2),\nonumber\\
&&\frac{\Gamma(~h(H)\to f\bar{f}~)}{\Gamma_{SM}(~h(H)\to
f\bar{f}~)}\simeq \frac{\Gamma(~h(H)\to VV~)}{\Gamma_{SM}(~h(H)\to
VV~)}\simeq \frac{\Gamma(~h(H)\to gg~)}{\Gamma_{SM}(~h(H)\to
gg~)}\simeq c_0^2 ( s_0^2),\label{rwid}\eea where $V$ denotes
$W,~Z$. Compared to SM, in addition to the modified $ht\bar{t}$ and
$hWW$ couplings, the charged Higgs $H^\pm$ will alter the decays
$h\to \gamma\gamma$ and $h\to Z\gamma$ via the one-loop. The
corresponding expressiones are given in the Appendix A.

The Higgs boson $\gamma\gamma$, $ZZ^*$ and $Z\gamma$ rates of HTM0
normalized to the SM values are respectively defined as
 \bea
R_{h(H)}(\gamma\gamma)&=&\frac{\sigma\left(~pp\to
h(H)~\right)}{\sigma_{SM}\left(~pp\to
h(H)~\right)}\frac{Br\left(~h(H)\to\gamma\gamma~\right)}{Br_{SM}\left(~h(H)\to\gamma\gamma~\right)}
\nonumber\\&\simeq&
c_0^2(s_0^2)\frac{\Gamma(~h(H)\to\gamma\gamma~)}{c_0^2(s_0^2)\Gamma_{SM}(~h(H)~)}
\frac{\Gamma_{SM}(~h(H)~)}{\Gamma_{SM}(~h(H)\to\gamma\gamma~)}
\simeq \frac{\Gamma(~h(H)\to\gamma\gamma~)}{\Gamma_{SM}(~h(H)\to\gamma\gamma~)},\nonumber\\
R_{h(H)}(ZZ^*)&=&\frac{\sigma\left(~pp\to
h(H)~\right)}{\sigma_{SM}\left(~pp\to
h(H)~\right)}\frac{Br\left(~h(H)\to
ZZ^*~\right)}{Br_{SM}\left(~h(H)\to ZZ^*~\right)}
\nonumber\\&\simeq&
c_0^2(s_0^2)\frac{c_0^2(s_0^2)\Gamma_{SM}(~h(H)\to
ZZ^*~)}{c_0^2(s_0^2)\Gamma_{SM}(~h(H)~)}
\frac{\Gamma_{SM}(~h(H)~)}{\Gamma_{SM}(~h(H)\to ZZ^*~)}
\simeq c_0^2(s_0^2),\nonumber\\
R_{h(H)}(Z\gamma)&=&\frac{\sigma\left(~pp\to
h(H)~\right)}{\sigma_{SM}\left(~pp\to
h(H)~\right)}\frac{Br\left(~h(H)\to
Z\gamma~\right)}{Br_{SM}\left(~h(H)\to Z\gamma~\right)}
\nonumber\\&\simeq& c_0^2(s_0^2)\frac{\Gamma(~h(H)\to
Z\gamma~)}{c_0^2(s_0^2)\Gamma_{SM}(~h(H)~)}
\frac{\Gamma_{SM}(~h(H)~)}{\Gamma_{SM}(~h(H)\to Z\gamma~)} \simeq
\frac{\Gamma(~h(H)\to Z\gamma~)}{\Gamma_{SM}(~h(H)\to
Z\gamma~)}.\label{rdef}\eea Where $\sigma\left(~pp\to h(H)~\right)$
is the total cross section of Higgs boson. The analytic expressions
in Eq. (\ref{rwid}) and Eq. (\ref{rdef}) may help us understand the
Higgs production and decay well. In our numerical calculations, we
take code Hdecay to consider the relevant higher order QCD and
electroweak corrections \cite{hdecay}.

\subsection{Scenario I}
For the scenario I, the light Higgs $h$ is the observed boson. Since
the observed $ZZ^*$ rate is consistent with the SM value, $c_0$ can
not be too small. Also, it is important to make sure that the
production rate of $H$ is small enough not to be detected at the
LHC. Thus, to obtain a large $c_0$ and a small $s_0$, we require $C>
A$ (see Eq. (\ref{CA})).

\begin{figure}[tb]
 \epsfig{file=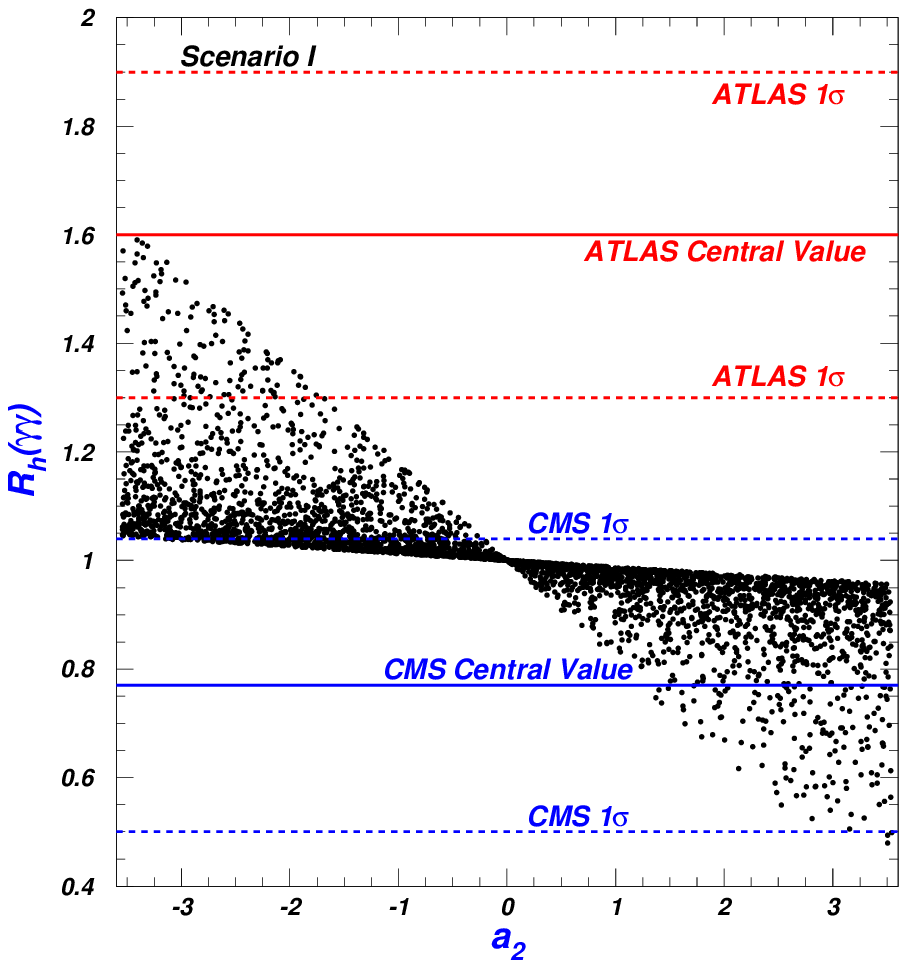,height=7.5cm}
  \epsfig{file=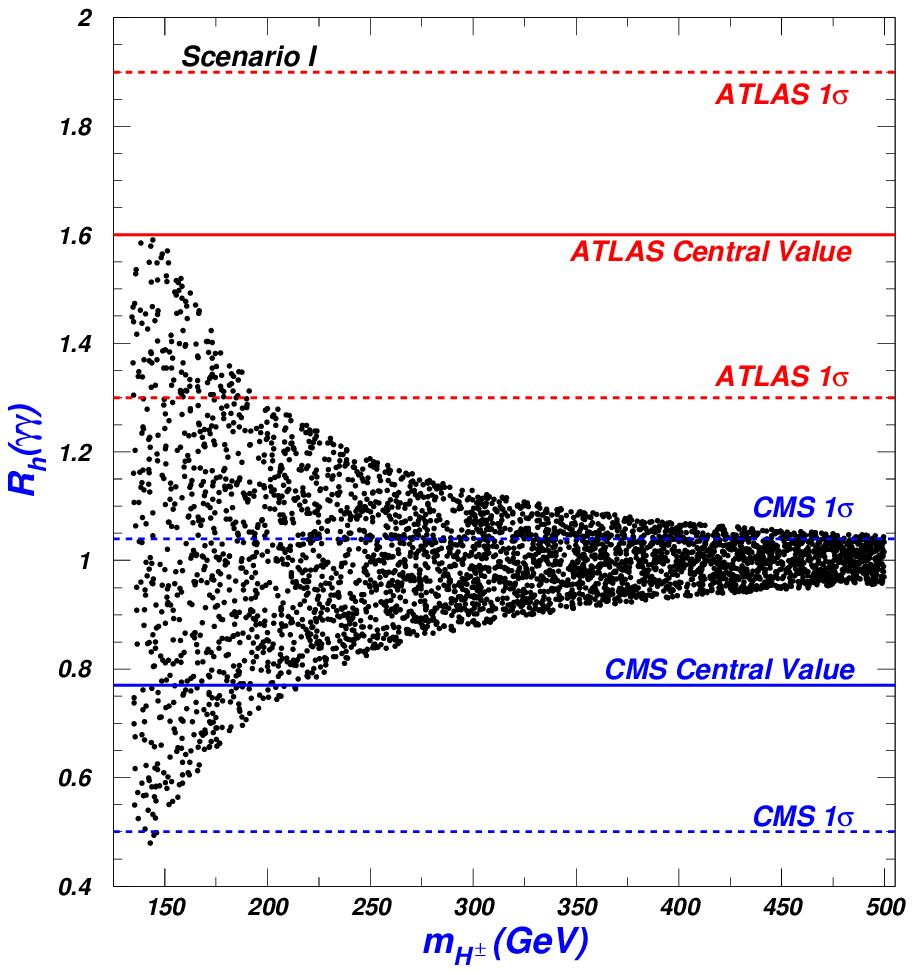,height=7.5cm}
\vspace{-0.4cm} \caption{The scatter plots of the parameter space
projected on the planes of $R_h(\gamma\gamma)$ versus $a_2$ and
$R_h(\gamma\gamma)$ versus $m_{H^{\pm}}$, respectively.} \label{i1}
\end{figure}

\begin{figure}[tb]
 \epsfig{file=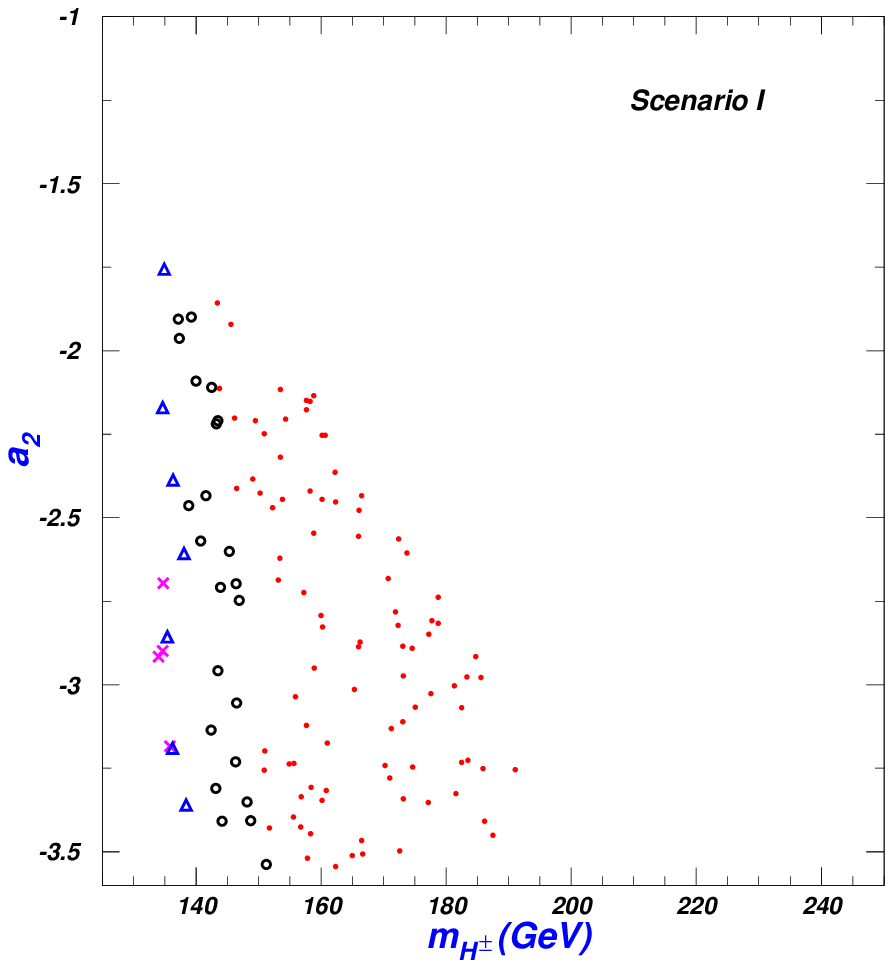,height=7.6cm}
  \epsfig{file=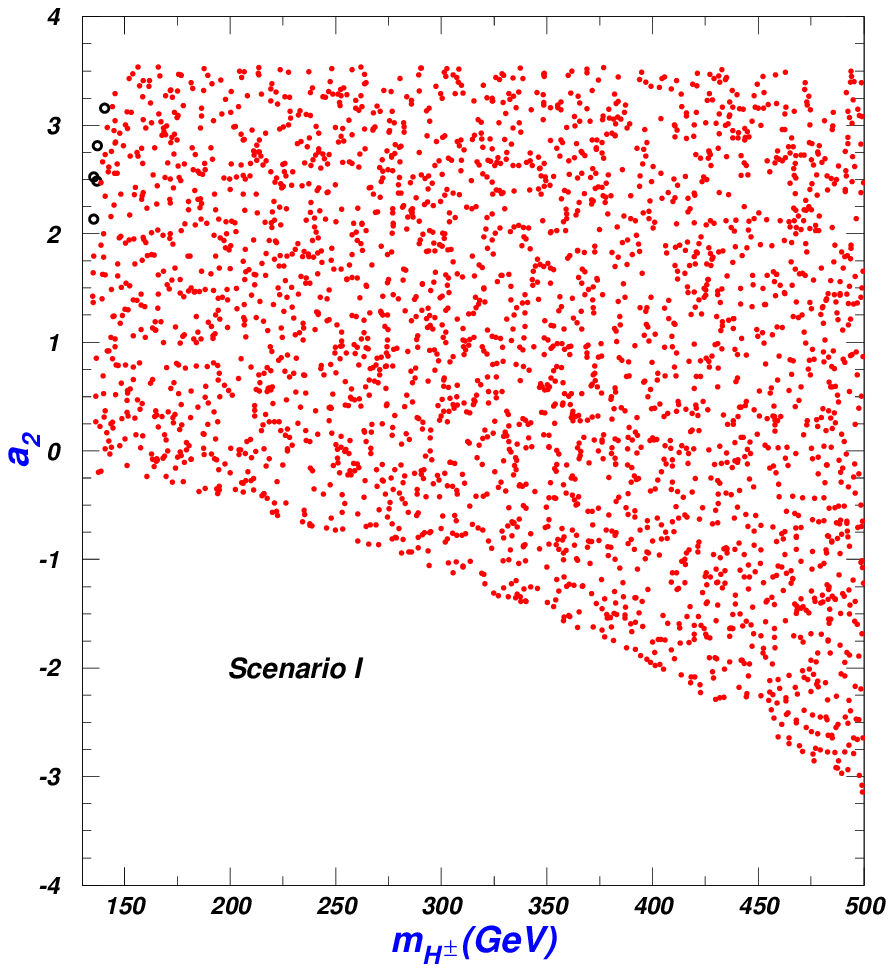,height=7.5cm}
\vspace{-0.4cm} \caption{The scatter plots projected on the plane of
$a_2$ versus $m_{H^{\pm}}$. For the left panel, $R_h(\gamma\gamma)$
is within $1\sigma$ range of ATLAS data. $0.86<c_0^2<0.90$ for the
crosses (pink), $0.90\leq c_0^2<0.95$ for the triangles (blue),
$0.95\leq c_0^2<0.98$ for the circles (black), and $0.98\leq
c_0^2<1.0$ for the bullets (red). The right panel is the same as the
left panel, but $R_h(\gamma\gamma)$ is within $1\sigma$ range of CMS
data.} \label{i2}
\end{figure}

\begin{figure}[tb]
 \epsfig{file=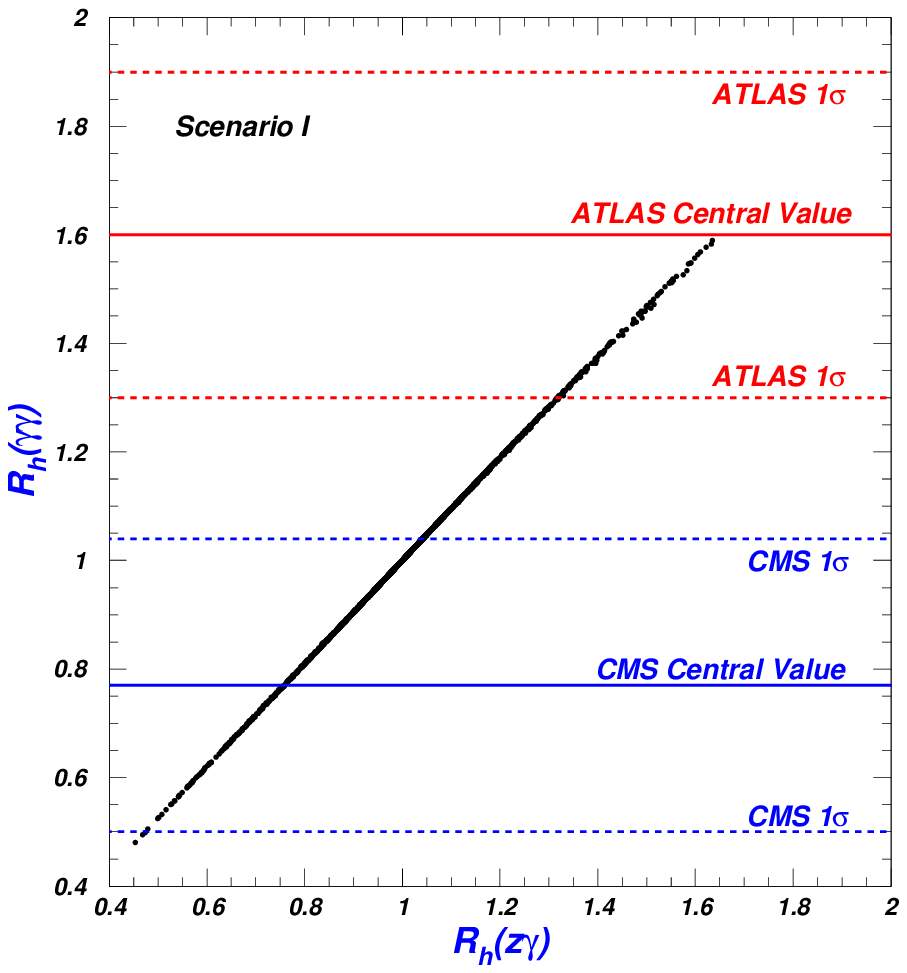,height=7.5cm}
\vspace{-0.4cm} \caption{The scatter plots of the parameter space
projected on the plane of $R_h(\gamma\gamma)$ versus
$R_h(Z\gamma)$.} \label{i3}
\end{figure}

In Fig. \ref{i1}, we plot $R_h(\gamma\gamma)$ versus $a_2$ and
$R_h(\gamma\gamma)$ versus $m_{H^{\pm}}$, respectively. The $h$
coupling to $H^+ H^-$ is sensitive to the parameter $a_2$, which
gives the additional contributions to the decay $h\to \gamma\gamma$
via one-loop. Fig. \ref{i1} shows that the $H^{\pm}$ contributions
to $R_h(\gamma\gamma)$ can interfere constructively with $W$
contributions for $a_2<0$ and interfere destructively for $a_2>0$,
leading $R_h(\gamma\gamma)>1$ and $R_h(\gamma\gamma)<1$, which are
respectively favored by the enhanced ATLAS diphoton data and the
suppressed CMS data. The magnitude becomes sizable as the increasing
of the absolute value of $a_2$ and the decreasing of $m_{H^{\pm}}$.

 In Fig. \ref{i2}, the samples with $R_h(\gamma\gamma)$ being within $1\sigma$
range of ATLAS and CMS diphoton data are projected on the plane of
$a_2$ and $m_{H^{\pm}}$. The left panel shows that the $1\sigma$
ATLAS diphoton data favors $-3.6<a_2<-1.8$ and $m_{H^{\pm}}<190$
GeV. While the CMS data favors $a_2>0$ and allow $a_2$ to be smaller
than 0 for enough large $m_{H^{\pm}}$. The left panel shows that,
for $R_h(\gamma\gamma)$ is within $1\sigma$ range of ATLAS diphoton
data, the samples lie in the region of $c_0^2 > 0.86$, and the vast
majority of them congregate the region of $c_0^2> 0.96$. The large
$m_{H^{\pm}}$ favors a large $c_0^2$. From the right panel, the
value of $c_0^2$ is larger than 0.98 for $R_h(\gamma\gamma)$ is
within $1\sigma$ range of CMS diphoton data. Due to $R_h(ZZ^*)\simeq
c_0^2$ (see Eq. \ref{rdef}), the inclusive $ZZ^*$ rate is outside
$1\sigma$ range of ATLAS data ($1.5\pm0.4$), but within $1\sigma$
range of CMS data ($0.92\pm0.28$). Besides, for such large $c_0^2$,
the corresponding $s_0^2$ is smaller than 0.14, which will suppress
the production rates of $H$ at the LHC sizably (see Eqs.
(\ref{rwid}) and (\ref{rdef})), leading that $H$ is not detected at
the LHC.

Fig. \ref{i3} shows $R_h(\gamma\gamma)$ versus $R_h(Z\gamma)$. We
find that the two rates are positively correlated, and the behavior
of $R_h(Z\gamma)$ is similar to that of $R_h(\gamma\gamma)$.
Further, the prediction of $R_h(Z\gamma)$ equals to that of
$R_h(\gamma\gamma)$ approximately.

\subsection{Scenario II}
For the scenario II, the heavy Higgs $H$ is the observed boson. The
parameter $s_0$ can not be very small to make the observed $ZZ^*$
rate to be consistent with the experimental data. Besides, it is
important to make sure that the production rate of $h$ is small
enough not to be detected at the LEP. Thus, we require $C < A$ to
obtain a large $s_0$ and a small $c_0$, (see Eq. (\ref{CA})).

\begin{figure}[tb]
 \epsfig{file=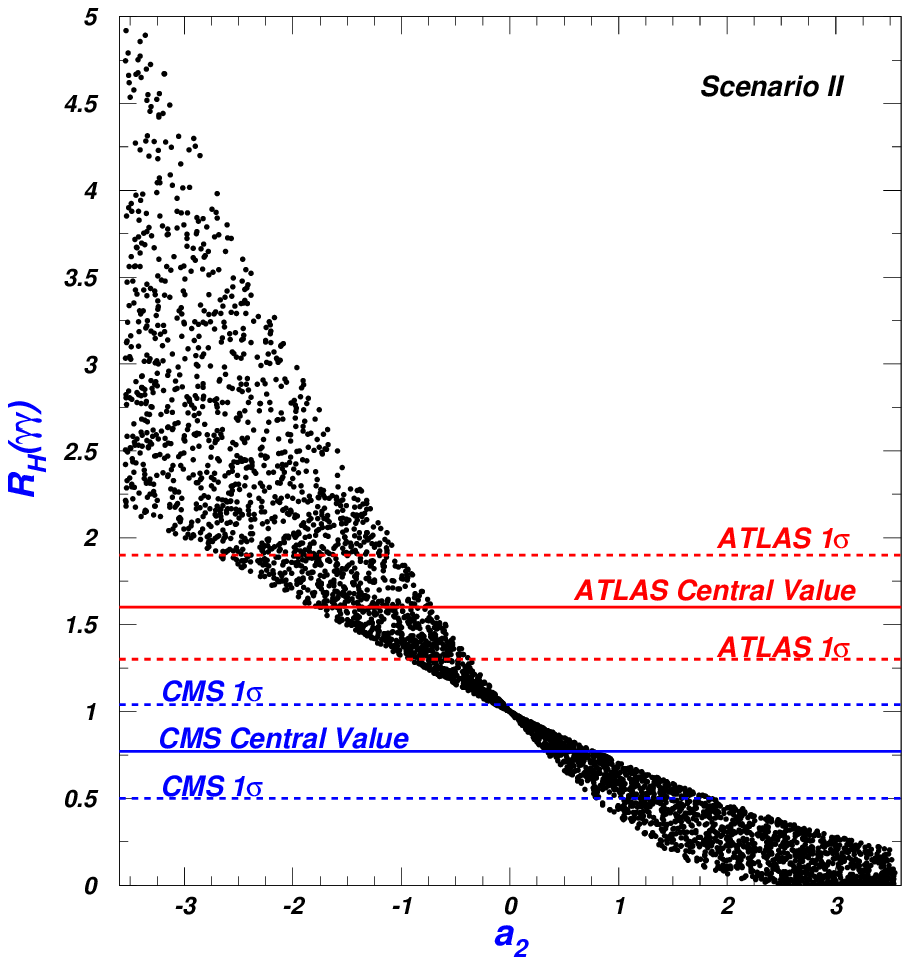,height=7.5cm}
  \epsfig{file=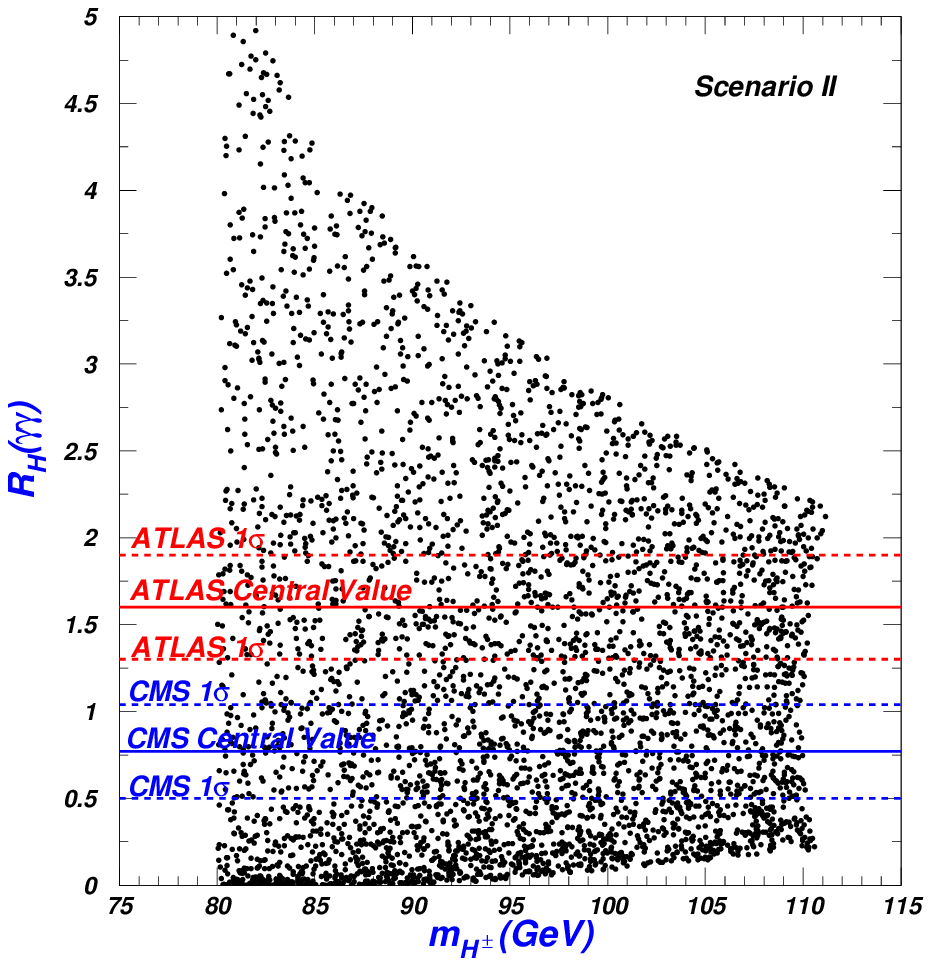,height=7.5cm}
\vspace{-0.4cm} \caption{Same as Fig. \ref{i1}, but for
$R_H(\gamma\gamma)$.} \label{ii1}
\end{figure}

\begin{figure}[tb]
 \epsfig{file=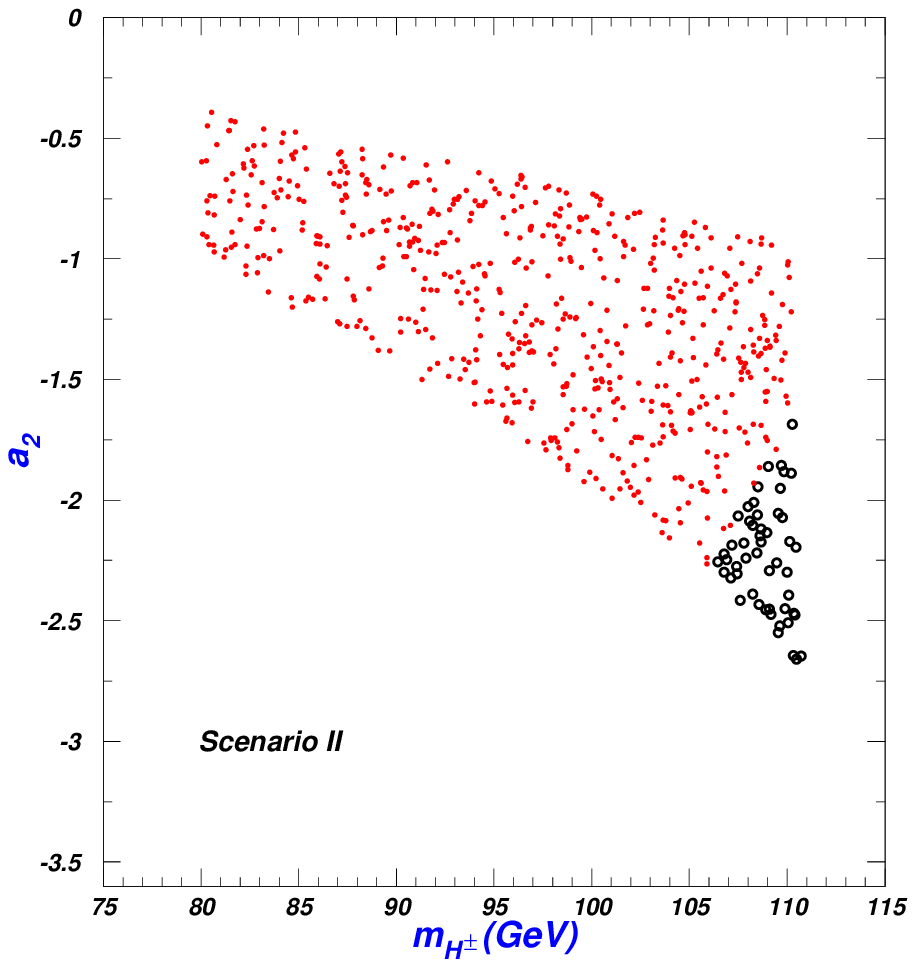,height=7.5cm}
  \epsfig{file=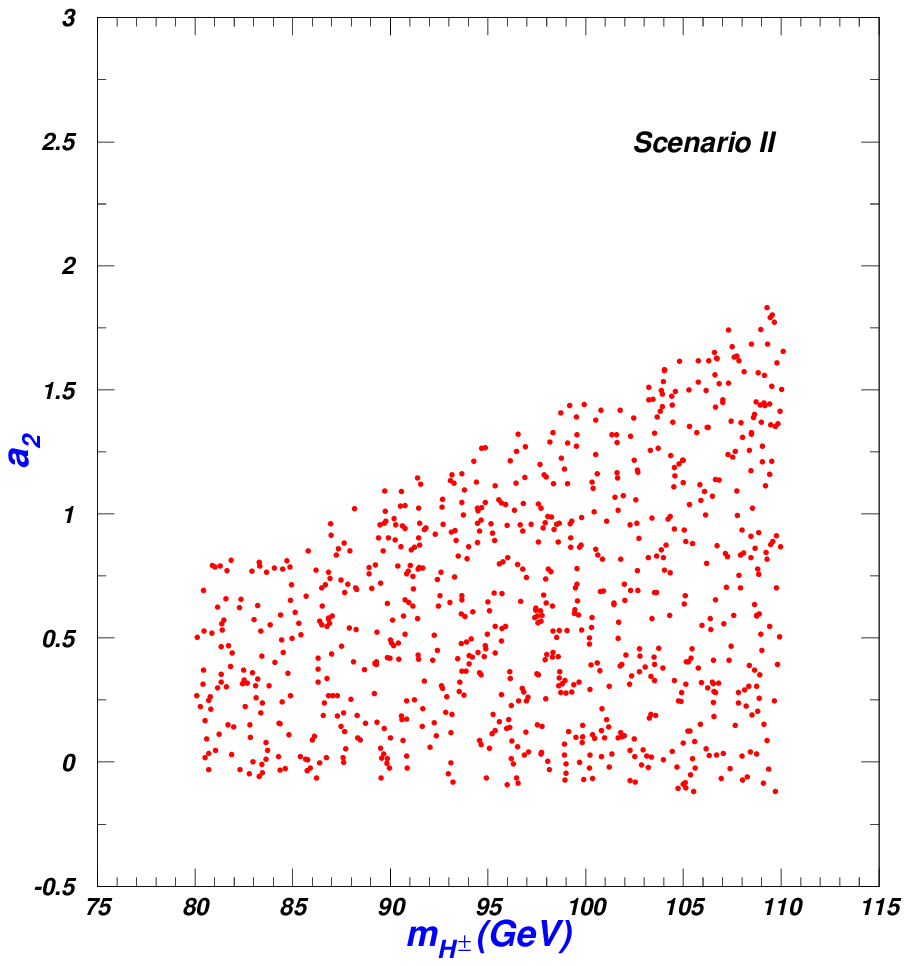,height=7.5cm}
\vspace{-0.4cm} \caption{The Scatter plots projected on the plane of
$a_2$ versus $m_{H^{\pm}}$. For the left panel, $R_H(\gamma\gamma)$
is within $1\sigma$ range of ATLAS data. $0.95\leq s_0^2<0.98$ for
the circles (black), and $0.98\leq s_0^2<1.0$ for the bullets (red).
The right panel is the same as the left panel, but
$R_H(\gamma\gamma)$ is within $1\sigma$ range of CMS data.}
\label{ii2}
\end{figure}

\begin{figure}[tb]
 \epsfig{file=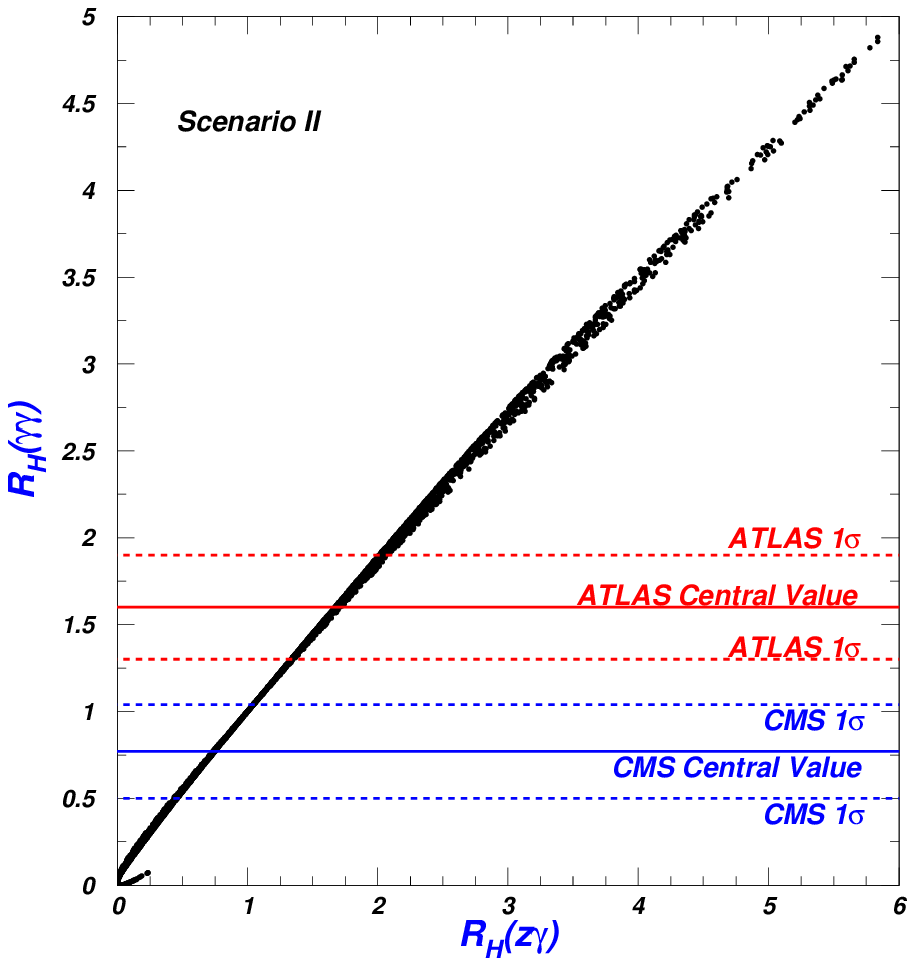,height=7.5cm}
\vspace{-0.4cm} \caption{Same as Fig. \ref{i3}, but for
$R_H(\gamma\gamma)$ versus $R_H(Z\gamma)$.} \label{ii3}
\end{figure}
In Fig. \ref{ii1}, we plot $R_H(\gamma\gamma)$ versus $a_2$ and
$R_H(\gamma\gamma)$ versus $m_{H^{\pm}}$, respectively. Similar to
$R_h(\gamma\gamma)$, $R_H(\gamma\gamma)$ is also larger than 1.0 for
$a_2 <0$ and smaller than 1.0 for $a_2>0$. $R_H(\gamma\gamma)$ can
reach 5.0 for $a_2\sim-3.5$ and $m_{H^{\pm}}\sim80$ GeV, which is
much larger than $R_h(\gamma\gamma)$ since $m_{H^{\pm}}$ for the
former is smaller than that for the latter.

In Fig. \ref{ii2}, the samples with $R_H(\gamma\gamma)$ being within
$1\sigma$ range of ATLAS and CMS diphoton data are projected on the
plane of $a_2$ and $m_{H^{\pm}}$. Fig. \ref{ii2} shows that
$-2.7<a_2<-0.4$ and $-0.1<a_2<1.9$ are respectively favored by the
$1\sigma$ ATLAS and CMS data. The left panel shows that, for
$R_H(\gamma\gamma)$ is within $1\sigma$ range of ATLAS diphoton
data, the samples lie in the region of $s_0^2 > 0.95$, and the vast
majority of them congregate the region of $s_0^2 > 0.98$. From the
right panel, the value of $s_0^2$ is larger than 0.98 for
$R_H(\gamma\gamma)$ is within $1\sigma$ range of CMS diphoton data.
The small $m_{H^{\pm}}$ favors a large $s_0^2$. Due to
$R_H(ZZ^*)\simeq s_0^2$ (see Eq. \ref{rdef}), the inclusive $ZZ^*$
rate is outside $1\sigma$ range of ATLAS data, but within $1\sigma$
range of CMS data. Besides, for such large $s_0^2$, the
corresponding $c_0^2$ is smaller than 0.05, and the cross section of
$e^+e^-\to Zh$ is below the upper limit presented by the LEP
\cite{lepn}.

In Fig. \ref{ii3}, we plot $R_H(\gamma\gamma)$ versus
$R_H(Z\gamma)$. Similar to scenario I, the two rates are also
positively correlated. Especially for the region favored by the
$1\sigma$ range of ATLAS and CMS data, the prediction of
$R_H(Z\gamma)$ equals to that of $R_H(\gamma\gamma)$ approximately.

\subsection{Scenario III}
For the scenario III, the observed signal is from the almost
degenerate $h$ and $H$. We assume that the mass splitting of $h$ and
$H$ is small enough not to be resolve at current statistics, but
large enough so that there is hardly interference between the
amplitudes of $h$ and $H$, $|m_H -m_h|\gg \Gamma(h),~\Gamma(H)$
\cite{dege}. Therefore, according to Eq. (\ref{hHmass}), both the
absolute values of $A-C$ and $B$ must be very small, but not to
equal to zero exactly. For this case, we can obtain a relation of
$m_{H^{\pm}}\simeq m_h\simeq m_H$ according to Eqs.
(\ref{matrix}),~(\ref{mcharge}) and (\ref{hHmass}).

\begin{figure}[tb]
 \epsfig{file=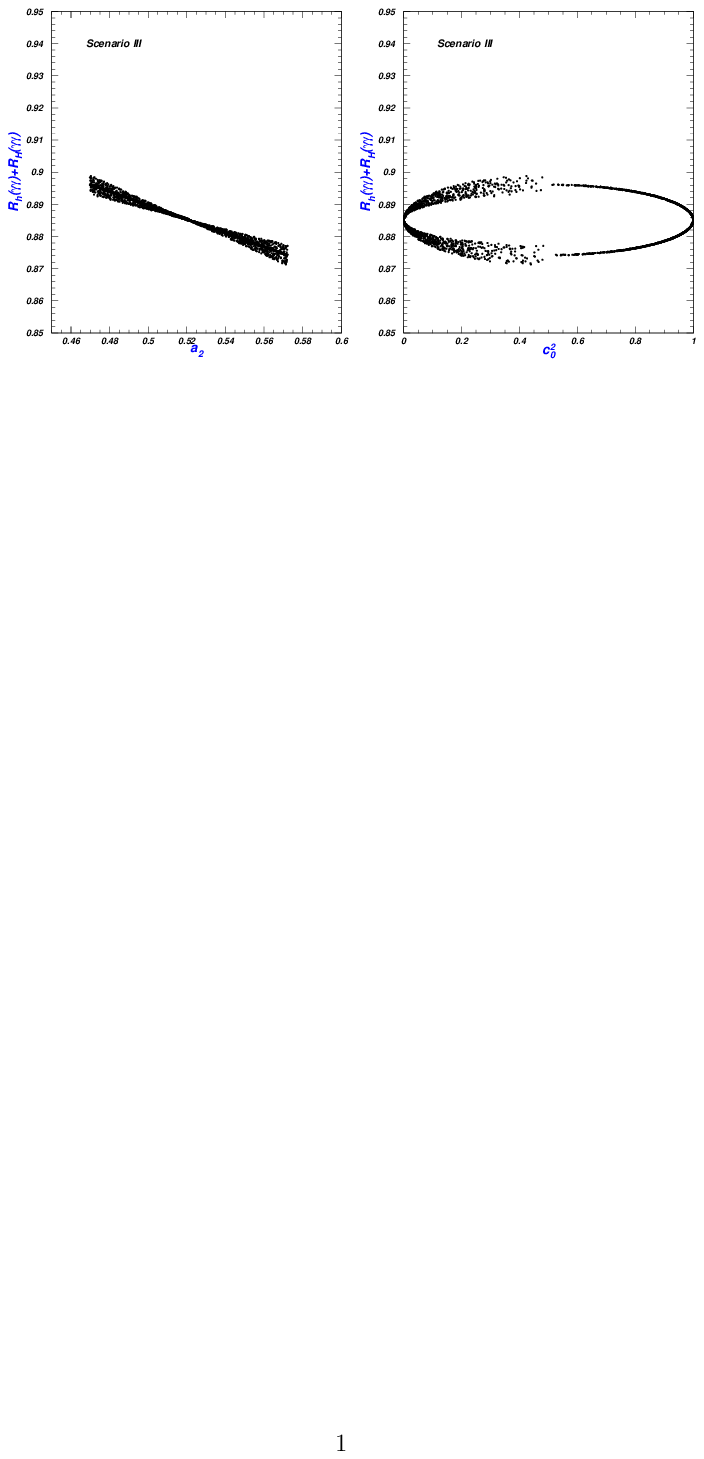,height=6.5cm}
\vspace{-0.4cm} \caption{The scatter plots of the parameter space
projected on the planes of $R_h(\gamma\gamma)+R_H(\gamma\gamma)$
versus $a_2$ and $c_0^2$, respectively.} \label{iii1}
\end{figure}

\begin{figure}[tb]
 \epsfig{file=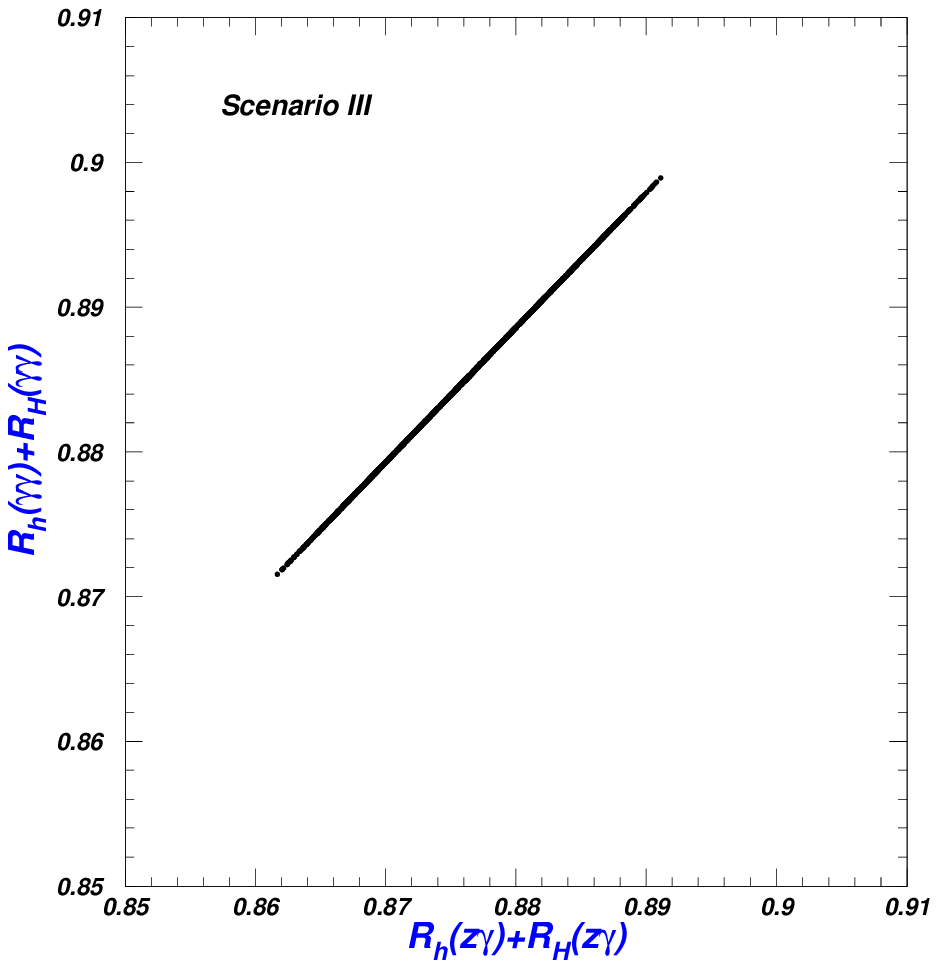,height=7.5cm}
\vspace{-0.4cm} \caption{Same as Fig. \ref{i3}, but for
$R_h(\gamma\gamma)+R_H(\gamma\gamma)$ versus
$R_h(Z\gamma)+R_H(Z\gamma)$.} \label{iii2}
\end{figure}

In Fig. \ref{iii1}, we plot $R_h(\gamma\gamma)$+$R_H(\gamma\gamma)$
versus  $a_2$ and $c_0^2$, respectively. We find that the Higgs
diphoton rate is suppressed compared to SM value,
$0.87<R_h(\gamma\gamma)$+$R_H(\gamma\gamma)<0.9$, which is outside
$1\sigma$ range of ATLAS diphoton data, but within $1\sigma$ range
of CMS diphoton data. Due to $a_1
>0$, $a_2$ must be larger than zero to obtain a very small $\mid
B\mid$ ($B=- a_1 v_d / 2 + a_2 v_d v_t$). Thus,
$R_h(\gamma\gamma)$+$R_H(\gamma\gamma)$ is smaller than 1.0 since
the $H^{\pm}$ contributions will interfere destructively with the
$W$ contributions for $a_2>0$. The right panel shows that the large
mixing angle $\theta_0$ may appear. The reason is that $\mid
A-C\mid$ still may be much smaller than $\mid B\mid$ although $\mid
B\mid$ is very small. Due to $R_h(ZZ^*)\simeq c_0^2$ and
$R_H(ZZ^*)=s_0^2$ (see Eq. (\ref{rdef})), the inclusive $ZZ^*$ rate
equals to SM prediction value approximately.

In Fig. \ref{iii2}, we plot $R_h(\gamma\gamma)+R_H(\gamma\gamma)$
versus $R_h(Z\gamma)+R_H(Z\gamma)$. We find that the two rates are
also positively correlated, and the correlation is more strong than
that of scenario II. $R_h(Z\gamma)+R_H(Z\gamma)$ is allowed to vary
in the narrow region, $0.86<R_h(Z\gamma)+R_H(Z\gamma)<0.89$.

\section{Conclusion}

In the Higgs triplet model with Y=0, we study the Higgs boson
$\gamma\gamma$ and $Z\gamma$ rates at the LHC. We studied three
different scenarios: (i) the observed boson is the light Higgs boson
$h$; (ii) it is the heavy Higgs boson $H$; (iii) the observed signal
is from the almost degenerate $h$ and $H$. We found that, for the
first two scenarios, the inclusive Higgs diphoton rates can be
enhanced or suppressed compared to the SM value, which is
respectively within $1\sigma$ range of ATLAS and CMS data. For the
scenario I, the ATLAS data favors $-3.6<a_2<-1.8$ and $m_{H^{\pm}}<$
190 GeV. The CMS data favors $a_2>0$ and allow $a_2$ to be smaller
than 0 for enough large $m_{H^{\pm}}$. For the scenario II, the
ATLAS and CMS diphoton data favor $-2.7<a_2<-0.4$ and
$-0.1<a_2<1.9$, respectively. For the first two scenarios, the
inclusive $ZZ^*$ rates are suppressed, which are outside $1\sigma$
range of ATLAS data and within $1\sigma$ range of CMS data. For the
third scenario, the Higgs diphoton rate is suppressed, which is
outside $1\sigma$ range of ATLAS data, and the $ZZ^*$ rate equals to
SM value approximately. Besides, the two rates of $h\to
\gamma\gamma$ and $h\to Z\gamma$ are positively correlated, and they
are approximately equal within the $1\sigma$ range of ATLAS and CMS
diphoton data.

\section*{Acknowledgment}
This work was supported by the National Natural Science Foundation
of China (NNSFC) under grant Nos. 11105116, 11005089, and 11175151.

\appendix
\section{The expressions for $\Gamma(h\rightarrow \gamma\gamma)$
and $\Gamma(h\rightarrow{Z\gamma})$} The charged fermion ($f$),
gauge boson ($W$) and scalar ($s$) can contribute to the decay
widths of $h\rightarrow \gamma\gamma$ and $h\rightarrow Z\gamma$,
which are given by \cite{geng,decayrr}
\begin{eqnarray}
\Gamma(h\rightarrow \gamma\gamma)&=&{\alpha^2 m_h^3\over 256\pi^3
v^2}\left|\sum_{f}N^c_fQ_f^2y_f
A^{\gamma\gamma}_{1/2}(\tau_f)+y_WA^{\gamma\gamma}_1(\tau_W)
+Q_s^2{v\mu_{hss^*}\over
2m_{s}^2}A^{\gamma\gamma}_0(\tau_{s})\right|^2\,,
\label{generalformulae1}\\
\Gamma(h\rightarrow{Z\gamma})&=&{\alpha^2 m_h^3\over
128\pi^3s_W^2c_W^2v^2}\left(1-m_Z^2/m_h^2\right)^3
\bigg|N_f^cQ_fy_f{(Q_R^Z+Q_L^Z)\over 2}A^{Z\gamma}_{1/2}(\tau_f,\lambda_f)\nonumber\\
&&+Q_WQ_W^Zy_W
A^{Z\gamma}_1(\tau_{W},\lambda_{W})+Q_sQ_s^Z{vg_{hss}\over2
m_{s}^2}A^{Z\gamma}_0(\tau_{s},\lambda_{s})
\bigg|^2\,,\label{generalformulae2}
\end{eqnarray}
where $\tau_i=m_h^2/4m_i^2$, $\lambda_i=m_Z^2/4m_i^2$, $Q_W=1$,
$Q_W^Z=c^2_W$. $Q_{f,s}$ are the electric charges of fermion and
scalar. $N_f^c$ is the color factor for fermion $f$.
$Q_{R,L(s)}^Z=I_{R,L(s)}^3-Q_{f(s)}s^2_W$ with $I_{R,L(s)}^3$ being
the third isospin components of chiral fermions (scalar). $y_f$ and
$y_W$ denote the Higgs couplings to $f\bar{f}$ and $WW$ normalized
to the corresponding SM values. $g_{hss}$ is the coupling constant
of $hss$. The loop functions $A^{\gamma\gamma}_{(0,\,1/2,\,1)}$ and
$A^{Z\gamma}_{(0,\,1/2,\,1)}$ in Eqs.~(\ref{generalformulae1}) and
(\ref{generalformulae2}) are defined as
\begin{eqnarray}
A^{\gamma\gamma}_{0}(\tau)&=&-[\tau-f(\tau)]\tau^{-2}\,,
A^{\gamma\gamma}_{1/2}(\tau)=2[\tau+(\tau-1)f(\tau)]\tau^{-2}\,,\,\nonumber\\
A^{\gamma\gamma}_{1}(\tau)&=&-[2\tau^2+3\tau+3(2\tau-1)f(\tau)]\tau^{-2}\,,\nonumber\\
A^{Z\gamma}_0(\tau,\lambda)&=&I_1(\tau,\lambda)\,,\;
A^{Z\gamma}_{1/2}(\tau,\lambda)=-2[I_1(\tau,\lambda)-I_2(\tau,\lambda)], \nonumber\\
A^{Z\gamma}_1(\tau,\lambda)&=&[2(1+2\tau)(1-\lambda)+(1-2\tau)]I_1(\tau,\lambda)-8(1-\lambda)I_2(\tau,\lambda)\,,
\end{eqnarray}
where
\begin{eqnarray}
I_1(\tau,\lambda)&=&-{1\over (\tau-\lambda)}+{1\over (\tau-\lambda)^2}[f(\tau)-f(\lambda)]+{2\lambda\over (\tau-\lambda)^2}[g(\tau)-g(\lambda)]\;,\nonumber\\
I_2(\tau,\lambda)&=&{1\over(\tau-\lambda)}[f(\tau)-f(\lambda)]\;,
\end{eqnarray}
with the functions $f(\tau)$ and $g(\tau)$ given by
\begin{eqnarray}
f(\tau)= \bigg\{\begin{array}{ll}
(\sin^{-1}\sqrt{\tau})^2\,,\hspace{60pt}& \tau\leq 1\\
-{1\over4}[\log{1+\sqrt{1-\tau^{-1}}\over1-\sqrt{1-\tau^{-1}}}-i\pi]^2\,,\quad
&\tau>1
\end{array}\;,\;
g(\tau)= \bigg\{\begin{array}{ll}
\sqrt{\tau^{-1}-1}(\sin^{-1}\sqrt{\tau})\,,\hspace{60pt}& \tau\leq 1\\
{\sqrt{1-\tau^{-1}}\over
2}[\log{1+\sqrt{1-\tau^{-1}}\over1-\sqrt{1-\tau^{-1}}}-i\pi]\,,\quad
&\tau>1\;.
\end{array}\nonumber\\
\end{eqnarray}

\section{The vacuum expectation values}
The minimization conditions for the tree-level Higgs potential are
\begin{eqnarray}
\left( - \mu^2 +  \lambda_0  v_d^2  -  \frac{a_1  v_t }{2} +
\frac{a_2  v_t^2 }{2} \right)  v_d &=& 0 \ , \label{min1}
\\
- M_{\Sigma}^2  v_t  +  b_4 v_t^3  -  \frac{a_1  v_d^2}{4 } +
\frac{a_2  v_d^2  v_t}{2} & = & 0 \ . \label{min2}
\end{eqnarray}
Solving the Eqs. (\ref{min1}) and (\ref{min2}) with
\textsf{Mathematica}, we can obtain the expressions of $v_t$ and
$v_d$ in terms of the Lagrangian parameters. However, their
expressions are very complicated and lengthy. Therefore, we assume
$v_t$ to be much smaller than 1, and give the approximate solutions
for $a_2\mu^2\geq 2M_\Sigma^2\lambda_{0}$, \bea
v_t&=&\frac{1}{a_1}\left(-\mu^2+\frac{a_1^2}{4a_2}+\frac{2 M_{\Sigma
}^2 \lambda_0}{a_2}+\frac{\sqrt{-128 \mu ^2 a_2 M_{\Sigma}^2
\lambda_0+(a_1^2+4 \mu^2 a_2+8 M_{\Sigma}^2
\lambda_0)^2}}{4a_2}\right), \label{vt1}\\
v_d&=&\sqrt{\frac{M_{\Sigma}^2}{a_2}+\frac{\mu^2}{2
\lambda_0}+\frac{a_1^2}{8 a_2 \lambda_0}+\frac{\sqrt{-128 \mu^2 a_2
M_{\Sigma }^2 \lambda_0+(a_1^2+4 \mu^2 a_2+8 M_{\Sigma}^2
\lambda_0)^2}}{8 a_2 \lambda_0}},
 \eea
and for $a_2\mu^2\leq 2M_\Sigma^2\lambda_{0}$, \bea
v_t&=&\frac{1}{a_1}\left(-\mu^2+\frac{a_1^2}{4a_2}+\frac{2 M_{\Sigma
}^2 \lambda_0}{a_2}-\frac{\sqrt{-128 \mu ^2 a_2 M_{\Sigma}^2
\lambda_0+(a_1^2+4 \mu^2 a_2+8 M_{\Sigma}^2
\lambda_0)^2}}{4a_2}\right), \label{vt2}\\
v_d&=&\sqrt{\frac{M_{\Sigma}^2}{a_2}+\frac{\mu^2}{2
\lambda_0}+\frac{a_1^2}{8 a_2 \lambda_0}-\frac{\sqrt{-128 \mu^2 a_2
M_{\Sigma }^2 \lambda_0+(a_1^2+4 \mu^2 a_2+8 M_{\Sigma}^2
\lambda_0)^2}}{8 a_2 \lambda_0}}.
 \eea
From Eqs. (\ref{vt1}) and (\ref{vt2}), $v_t$ approaches to 0 for
$a_1\rightarrow 0$, which is understandable since $a_1$ is the
coefficient of the only term in the Lagrangian breaking the
custodial symmetry.

\end{document}